\documentclass[prd, aps, twocolumn, superscriptaddress, floatfix, nofootinbib, amsmath, amssymb, preprintnumbers]{revtex4-2}
\usepackage{graphicx}
\usepackage{dcolumn}
\usepackage{verbatim}
\usepackage{bm}
\usepackage{lmodern}

\usepackage{tikz}
\usepackage{pgfplots}
\pgfplotsset{compat=newest,every axis plot/.append style={line width=1pt}}

\usepackage{tabularx}

\usepackage{hyperref}
\hypersetup{
colorlinks=true,
citecolor=blue,
linkcolor=blue,
urlcolor=blue}

\usepackage{enumitem}
\newenvironment{foo}
  {\obeylines\everypar{\textbullet\hspace{\labelsep}}}
  {}

\definecolor{lightgray}{gray}{0.9}
\definecolor{Amber}{rgb}{1.0, 0.75, 0.0}
\definecolor{blizzardblue}{rgb}{0.67, 0.9, 0.93}

\begin{document}

\title{A narrow-band parameterization for the stochastic gravitational wave background 
}

\author{Tianyi Xie}
\email[Email: ]{xietianyi@mail.ustc.edu.cn}
\affiliation{Deep Space Exploration Laboratory/School of Physical Sciences, University of Science and Technology of China, Hefei, Anhui 230026, China}
\affiliation{CAS Key Laboratory for Researches in Galaxies and Cosmology/Department of Astronomy, School of Astronomy and Space Science, University of Science and Technology of China, Hefei, Anhui 230026, China}

\author{Dongdong Zhang}
\email[Email: ]{don@mail.ustc.edu.cn}
\affiliation{Deep Space Exploration Laboratory/School of Physical Sciences, University of Science and Technology of China, Hefei, Anhui 230026, China}
\affiliation{CAS Key Laboratory for Researches in Galaxies and Cosmology/Department of Astronomy, School of Astronomy and Space Science, University of Science and Technology of China, Hefei, Anhui 230026, China}
\affiliation{Kavli IPMU (WPI), UTIAS, The University of Tokyo, Kashiwa, Chiba 277-8583, Japan}

\author{Jie Jiang}
\email[Email: ]{jiejiang@pusan.ac.kr}
\affiliation{CAS Key Laboratory for Researches in Galaxies and Cosmology/Department of Astronomy, School of Astronomy and Space Science, University of Science and Technology of China, Hefei, Anhui 230026, China}
\affiliation{Center for Cosmological Constant Problem, Pusan National University, Busan 46241, Republic of Korea}
\affiliation{Department of Physics, Pusan National University, Busan 46241, Republic of Korea}

\author{Jia-Rui Li}
\email[Email: ]{jr981025@mail.ustc.edu.cn}
\affiliation{Deep Space Exploration Laboratory/School of Physical Sciences, University of Science and Technology of China, Hefei, Anhui 230026, China}
\affiliation{CAS Key Laboratory for Researches in Galaxies and Cosmology/Department of Astronomy, School of Astronomy and Space Science, University of Science and Technology of China, Hefei, Anhui 230026, China}

\author{Bo Wang}
\email[Email: ]{ymwangbo@ustc.edu.cn}
\affiliation{Deep Space Exploration Laboratory/School of Physical Sciences, University of Science and Technology of China, Hefei, Anhui 230026, China}
\affiliation{CAS Key Laboratory for Researches in Galaxies and Cosmology/Department of Astronomy, School of Astronomy and Space Science, University of Science and Technology of China, Hefei, Anhui 230026, China}

\author{Yi-Fu Cai}
\email[Email: ]{yifucai@ustc.edu.cn}
\affiliation{Deep Space Exploration Laboratory/School of Physical Sciences, University of Science and Technology of China, Hefei, Anhui 230026, China}
\affiliation{CAS Key Laboratory for Researches in Galaxies and Cosmology/Department of Astronomy, School of Astronomy and Space Science, University of Science and Technology of China, Hefei, Anhui 230026, China}

\begin{abstract}
In light of the non-perturbative resonance effects that may occur during inflation, we introduce a parametrization for the power spectrum of the stochastic gravitational wave background (SGWB) characterized by narrow-band amplification. We utilize the universal \(\Omega_\text{GW}\propto k^3\) infrared limit, applicable to a wide array of gravitational wave sources, to devise a robust yet straightforward parameterization optimized for Markov Chain Monte Carlo (MCMC) analyses. This parameterization is demonstrated through select examples where its application is pertinent, and we discuss the advantages of this approach over traditional parametrizations for narrow-band scenarios. To evaluate the sensitivity of our proposed model parameters, we apply a mock likelihood based on the CMB-Stage4 data. Furthermore, we explicate the computational process for the mapping relationship between the foundational model parameters and our parameterized framework, using a two-field inflation model that resonantly amplifies gravitational waves (GWs) as an example.

\end{abstract}

\maketitle


\section{\label{Sec1}Introduction}
The recent discovery of the evidence from Pulsar Timing Arrays \cite{NANO2306,ParkersPTA,EeuropeanPTA,ChinaPTA} supporting the existence of a stochastic gravitational wave background (SGWB) (see \cite{Christensen2019} for a review) has illuminated new aspects of GW observation. In contrast to GWs generated by specific single events, such as those observed in binary pulsar systems \cite{Smith2006} and stellar-mass black hole merger events \cite{LIGO}, the SGWB arises from a multitude of random, independent events. Various phenomena contribute to SGWB \cite{Christensen2019}, including black hole/neutron star binaries \cite{7,8,9,10,11,12}, first-order phase transitions \cite{13,14,15,16,17,18,19,20}, spectator fields \cite{21,23,24,25,biagetti}, reheating/preheating after inflation \cite{26,27,29,30,31,32,33,34,35,36,37,38,39,40,41,42,Dufaux2007}, topological defects \cite{43,44,45,46,47,48}, primordial magnetic fields \cite{49,50,51,52,53,54,55,56}, and primordial perturbations from inflation. Many of these events trace back to the primordial era of the universe, produced as primordial gravitational waves (PGWs) \cite{Carr1980,Rev2010}.
 
Though cosmological models possess unique details, their observable characteristics are often constrained by the limitations of current observational capabilities, exhibiting similarities. Consequently, parameterization methods for the SGWB power spectrum have been devised and assessed against observational data. Among these, the ``broken power-law" parameterization, which assumes a power-law behavior for both the ultraviolet (UV) and infrared (IR) limits, is widely accepted. It is applicable to a variety of models that generate SGWB, such as phase transitions \cite{PT0, PT1, PT2}, inflationary models \cite{USR, DIP, Cai2019}, and scalar-induced gravitational waves \cite{Friction}. Another prevalent approach is the ``log-normal" parameterization; it employs a normal distribution function in the logarithm of the wave number \(k\). This method is useful for representing distributions across an extensive range of parameter spaces, substituting \(k\) with other variables. Examples include scalar-induced gravitational waves \cite{pi2020, Chen2022}, the distribution of reionization bubble sizes \cite{Dvorkin2009}, or the mass distribution of primordial black holes \cite{Gong2018}. Both parameterization schemes have been applied to interpret the NANOGrav signals \cite{NANO,Afzal2023}.

Nevertheless, there are scenarios where these parameterizations are insufficient. For instance, during the inflationary epoch, a non-perturbative resonance can trigger a narrow-band amplification of the primordial tensor power spectrum. This allows for a reduced tensor-to-scalar ratio, which in turn mitigates the lower limit on the field excursion established by previous research \cite{lyth1996}. Under such conditions, the two popular parameterizations fail to capture the dynamics accurately. 
In this work, we introduce a new parameterization that accurately characterizes the SGWB generated by a two-field inflation model that resonantly amplifies GWs. We demonstrate its superior performance and the necessity for its adoption over conventional parameterizations. In addition, by analyzing the specific traits of narrow-band GW production, particularly with scalar-induced gravitational waves (SIGW) as an example, we present evidence supporting the broad applicability of our parameterization in various ``narrow-band'' SGWB scenarios. Furthermore, we establish that our proposed parameterization is valid for narrow-band tensor perturbation amplification mechanisms during both the radiation-dominated era and the inflationary period.

Beyond the congruence of our parameterization with numerical results, we investigate the anisotropy of the cosmic microwave background (CMB) B-mode polarization as a diagnostic tool. Utilizing Monte Carlo Markov Chain (MCMC) analysis on mock data, our findings underscore the practicality of implementing our parameterization in constraining the parameter space using CMB observations. This is exemplified with mock data geared towards the forthcoming CMB-S4 experiment \cite{Collaboration2022}.
The organization of this article is as follows: Section \ref{Sec2} introduces our parameterization, addressing its suitability and applicability for narrow-band GW production. Section \ref{Sec3} details a comparative analysis of our parameterization with conventional parameterizations by using the two-field resonance model, highlighting the benefits of our approach in modeling narrow-band GW production. In Section \ref{Sec4}, we describe our settings of the MCMC simulations and the forecasts applied to our parameterization based on CMS-S4 mock data. We conclude with a discussion in Section \ref{Sec5}.

\section{\label{Sec2}Parameterization}
The parameterization of the SGWB is chosen over reliance on specific model results for two primary reasons. The first is the limitation of observational capabilities in distinguishing between different models, which, despite theoretical variations in SGWB production, yield similar observational signals. This warrants a model-independent approach. Secondly, a characteristic \(k^3\) scaling in the IR limit for various regimes, particularly narrow-band SGWB, is common. 

According to Ref. \cite{Infrared}, this scaling is applicable under certain conditions:

\begin{enumerate}[label=\upshape(\Roman*)]
\item\label{condition1} \(k\) is smaller compared to all the scales associated with the source term, such as \(k(\eta_1-\eta_2)\ll1\) and \(|\mathbf{k-p}|\approx |-\mathbf{p}|\), where \(p\) is an integrated wavenumber index of the source and \(\eta_1,\eta_2\) are two moments when the source still exists.
\item\label{condition2} The energy-momentum tensor should possess a comparably general form.\begin{equation}\begin{aligned}T_{ab}(\tau,\mathbf{k})=v_a(\tau,\mathbf{k})v_b(\tau,\mathbf{k})+\sum_I\partial_a\phi_I(\tau,\mathbf{k})\partial_b\phi_I(\tau,\mathbf{k})
\\
I=1,2,\cdots, \text{ as different scalar fields}~ .\end{aligned}\label{gem} \end{equation}
\item\label{condition3} The integral over wavenumber for computing \(\Omega_{\rm GW}\) after taking \(k \rightarrow 0\) should be finite. Namely,
\begin{equation}
\begin{aligned}
0<\int d \ell\left[\left(2 \mathcal{P}_{v}+3 \mathcal{P}_{w}\right)^{2}+5 \mathcal{P}_{w}^{2}+4\sum_I \mathcal{P}_{\phi}^{2}\right]<\infty
\end{aligned}~ , \end{equation}
where, 
\begin{equation}\begin{aligned}
&\left\langle v^{a}\left(\boldsymbol{\ell},\tau_{1}\right) v^{c *}\left({\boldsymbol{q}},\tau_{2}\right)\right\rangle=\delta^{(3)}(\boldsymbol{\ell}-\boldsymbol{q}) \frac{2 \pi^{2}}{\ell^{3}} 
\\
&\times \ell^{2}\left[\mathcal{P}_{w}\left(\tau_{1}, \tau_{2}, l\right) \pi^{a c}(\ell)+\mathcal{P}_{v}\left(\tau_{1}, \tau_{2}, l\right) \hat{\ell}^{a} \hat{\ell}^{c}\right] 
\\
&\left\langle\phi_I\left(\boldsymbol{\ell},\tau_{1}\right) \phi_J^{*}\left(\boldsymbol{q},\tau_{2}\right)\right\rangle
\\
&=\delta_{IJ}\delta^{(3)}(\boldsymbol{\ell}-\boldsymbol{q}) \frac{2 \pi^{2}}{\ell^{3}} \mathcal{P}_{\phi_I}\left(\tau_{1}, \tau_{2}, \ell\right),
\end{aligned}~ , \end{equation}
in which \(\pi^{a c}(\ell)=\delta^{a b}-\hat{\ell}^{a} \hat{\ell}^{c}\) and \(\mathcal{P}_{w}, \mathcal{P}_{v}\) are respectively longitudinal and perpendicular part of the power spectrum of \(\langle vv\rangle\), while \(\mathcal{P}_{\phi_I}\) represents the power spectrum of the scalar field noted by \(I\). We have assumed the two-point function between different scalar fields should be zero. 
\item\label{condition4} Modes of interest reenter the Hubble horizon during the radiation-dominated era to produce GW (or GW is produced during the inflationary era). 
\end{enumerate}

With a transient source that is spiky on wavenumber spectrum, its GW production generally is also narrow-band. For a spiky source, the integral for GW production in \ref{condition3} tends to be finite thus generating a \(\propto k^3\) IR limit. For a transient source, the \(k \rightarrow 0\) limit for \ref{condition1} is easier to achieve. That means narrow-band GW production tends to indicate a \(k^3\) scaling on the IR side. 
On the other hand, when the GW production band is narrow enough, we are required to describe UV limit by an exponential cutoff. 
As a result, we are motivated to propose a parametrization of \(\Omega_{\rm GW}\) when the amplification band is narrow, which provides a practical approach to SGWB signal discovery. 

As a preliminary step, our focus lies on the scale-dependent part of GW production. Considering the \(\propto k^3\) scaling mentioned above, the IR side of the peak remains fixed. On the UV side, we apply a simple exponential cutoff, instead of the typical power-law choice. We take the view of drawing with a log-log axis, so in our context ``exponential function" refers to that w.r.t. \(\ln k\), and power law is expressed with a ``linear function". We then propose the parametrization of the power spectrum as the combination of linear and exponential functions: 
\begin{equation}
\ln f(k)\sim 3\ln k-\exp(g\ln k)\label{liexp} ~ ,
\end{equation}
with $g$ a new dimensionless parameter.

Our approach has the following unique aspects compared to previous studies such as Ref. \cite{pi2020} and \cite{NANO}, making it most suitable for constraining the parameter space using statistical methods such as MCMC.

Firstly, we superpose an exponential term directly onto a linear term. This contrasts with adding a cutoff onto the log-normal function, namely a quadratic curve on a log-log axis. 
With no cutoff term \cite{pi2020,Zhou}, a log-normal function can only act as a rough approximation, lacking the power-law IR scaling and the asymmetry between the IR and UV sides. 
With a UV cutoff term on the log-normal function \cite{NANO}, it remains tricky even when we neglect the different scaling on the IR side. Two parameters are used to describe the quadratic term and one of them will degenerate with the parameter controlling the exponential term. It prevents the MCMC from yielding meaningful results.

Secondly, we apply the parameter controlling asymmetry inside the exponent. We use \(\ln f(k)\sim 3\ln k-\exp(g\ln k)\) rather than \(\ln f(k)\sim 3\ln k-g\exp(\ln k)\). This crucial deviation from the common case (e.g. Ref.\cite{NANO}) provides analytical convenience. For \(g \gg 1\), our choice screens out the contribution from the cutoff term to the IR side, while the other choice will enhance it. It stabilizes the parametrization and helps us treat the IR and UV sides separately: the linear term dominates the IR side, whereas the cutoff term dominates the UV side. For \(g\rightarrow 0\) the case is no longer narrow-band, where the other choice performs better. This is also shown in the discussion about Fig.\eqref{SIGW}.

Thirdly, it is interesting to draw an analogy between the parametrization and the scaling behavior of other physical systems. For instance, both the Planck black-body emission power spectrum and the heat capacity of crystals in a quantum scenario exhibit different behavior in the UV and IR limits, controlled by exponential functions. 

As a subsequent step, we append terms to \eqref{liexp}, independent of \(k\) and with no new parameters introduced, to ensure \(\frac{\partial}{\partial k}f(k)|_{k_*}=0\) and \(f(k)|_{k_*}=h_R\):
\begin{equation}
\begin{aligned}
f(k)=&h\exp\left(3 \ln\left(k/k_*\right)+\left(1-\exp\left(g \ln\left(k/k_*\right)\right)\right)\frac{3}{g}\right)
\end{aligned}
\label{para}
 \end{equation}
Here, \(h\) signifies the peak's height; \(k_*\) refers to the ultimate point's position; \(g\) represents the shape (the damping speed of the UV side), and the asymmetry. The 3 parameters \(h,g,k_*\) are largely independent, with only \(g\) influencing the peak's actual shape.

Moving forward, despite the \(k^3\) scaling in Ref.\cite{Infrared} was derived for GW production generally in radiation-dominated era, we can extend the conclusion to the inflationary era. Namely, for GW source fulfilling \ref{condition1}, \ref{condition2}, \ref{condition3}, there's also IR \(\propto k^3\) limit in inflationary GW production. 
The derivation is similar, with only displacing the green function and the integrated time interval. 
For the radiation-dominated era, the green function with a Heaviside step function \(\Theta\) reads 
\begin{equation}\begin{aligned}
&G(\tau,\tau')''+(k^2-\frac{a''}a)G(\tau,\tau')=\delta(\tau-\tau')
\\
&G(\tau,\tau')=\frac{\sin(k(\tau-\tau'))}{k}\Theta(\tau-\tau')
\end{aligned}~,\label{g1}\end{equation}
and the solution integral begins with conformal time \(\tau=0\)
\begin{equation}
h_{\mathbf{k},\lambda}=\frac{1}{a(\tau)}\int d\tau'a(\tau')G(\tau,\tau')S_{\mathbf{k},\lambda}(\tau')d\tau'.
\end{equation}
While for the inflationary era, the green function is 
\begin{equation}\begin{aligned}
&g(\tau,\tau')''+2\mathcal Hg(\tau,\tau')'+k^2g(\tau,\tau')=\delta(\tau-\tau')
\\
&g_k(\tau,\tau')=\frac{1}{k^3\tau'^2}\left[-k(\tau-\tau')\cos(k(\tau-\tau'))\right.
\\
&\quad\quad\quad\left.+\sin(k(\tau-\tau'))(1+k\tau\tau')    \right]\Theta(\tau-\tau')
\end{aligned}~ .\label{g2} \end{equation}
The solution integral is over negative value of conformal time
\begin{equation}\begin{aligned}
h_{\mathbf{k},\lambda}=&\int d\tau'g(\tau,\tau')S_{\mathbf{k},\lambda}(\tau')d\tau'.
\end{aligned}\end{equation}
In both cases, when taking the limit \(k \rightarrow 0\) the lowest order of green function under is proportional to \(k^0\), which ensures the reproduction of \(k^3\) scaling in inflationary era. 

In addition, since 
\begin{equation}
\begin{aligned}
\Omega_{\rm GW}(k,\tau)&=\frac{1}{12}\frac{k^2}{a^2 H^2} \Delta_t^2(k,\tau)~ 
\end{aligned}\label{proportion}
\end{equation}
inside horizon\cite{Boyle,Saito2010}, \(\Delta_t^2\) (primordial tensor power spectrum) will be proportional to \(\Omega_{\rm GW}\) (GW density power spectrum) thus have the same \(k\) dependence, because it is measured at horizon exit by definition and frozen outside horizon. As a result, we are guaranteed to impose the same parametrization on \(\Delta_t^2\).
In that case, we consider incorporating primordial contributions from single-field inflation which is scale-invariant. Given that the tilt \(n_t\) in the standard case is almost zero, it becomes overshadowed by the peak contribution within the amplified wavenumber range. Consequently, we parameterize as:
\begin{equation}
\begin{aligned}
\Delta_t^2=&(1+f(k))A_s r \left(\frac{k}{k_{\text{pivot}}}\right)^{n_t}
\\
=&\left[h_R\exp\left(3 \ln(\frac{k}{k_*})+(1-\exp(g \ln(\frac{k}{k_*})))\frac{3}{g}\right)+1\right]
\\
&\times A_s r \left(\frac{k}{k_{\text{pivot}}}\right)^{n_t}~ ,
\end{aligned}
\label{combine}
\end{equation}
where now \(h_R\) refers to peak's relative height w.r.t. the scale-invariant one. This reverts to the standard power-law spectrum when \(f(k)=0=h_R\).

Nonetheless, it's critical to note that the relative height \(h_R\) is not a physical quantity. The amplification mechanism is generally independent of the slow-roll regime, rendering \(h_R\) irrespective of the tensor-to-scalar ratio \(r\) or the slow-roll parameter \(\epsilon\) (\(r=16\epsilon=-8 n_t\) in the standard case). Thus, defining \(h_E:=h_R*r\) as the effective height, a physical parameter, is more suitable. In our following discussion, we explore how this functions in the MCMC process.

\section{\label{Sec3}Examples}
The strong instance supporting our parameterization and direct motivation to develop our parameterization is a two-field inflation model that resonantly amplifies GWs \cite{PRL,Zhou}. 

The GW production occurs during the inflationary era, and we will parameterize \(\Delta_t^2\) as mentioned in the previous section. This resonance model does not predict any extra curvature perturbation, so in this case, we preserve the power law of scalar power spectrum parameterization. This enables us to utilize CMB anisotropy to conduct MCMC as addressed in the next section, since otherwise, a scale-dependent curvature perturbation will contradict CMB observation. 

The generated primordial tensor power spectrum takes the form as described in \cite{Zhou}
\begin{equation}\begin{aligned}
&\Delta^2_{t}\left(k, \tau_{\text {end }}\right) 
=\frac{4}{\pi^{4} M_{p}^{4}} k^{3} \int_{0}^{\infty} d p p^{6} \int_{-1}^{1} d \cos \theta \sin ^{4} \theta 
\\
&\quad \times \mid \int_{\tau_{0}}^{\tau_{\text {end }}} d \tau_{1} g_{k}\left(\tau_{\text {end }}, \tau_{1}\right)
\\
&\quad
\left(\delta \phi_{p}\left(\tau_{1}\right) \delta \phi_{|\boldsymbol{k}-\boldsymbol{p}|}\left(\tau_{1}\right)\right.
\left.+\delta \chi_{p}\left(\tau_{1}\right) \delta \chi_{|\boldsymbol{k}-\boldsymbol{p}|}\left(\tau_{1}\right)\right)\left.\right|^{2}~ ,
\end{aligned}\label{Ph} \end{equation}
with the green function defined in \eqref{g2}.
\(\delta\phi_k(x),\delta\chi_k(x)\) are the perturbations of two scalar fields. \(\chi\) continues slow-rolling during inflation and dominates at the end, generating a normally scale-invariant scalar perturbation (and a small scale-invariant tensor perturbation, by model construction). Meanwhile, the amplified \(\phi\) serves as a source of tensor perturbation, enhancing the production of gravitational waves during the inflationary era. One may attempt to parameterize \(\delta\phi_k(x)\) to parameterize \(\Delta_t^2\), but it is model-dependent and introduces superfluous complexity evaluating the integral. Intense amplification of the perturbation amplifies the error for the analytic approximation as well, which could accumulate significantly after integration.

\begin{figure}
\centering
 \includegraphics[width = 240pt]{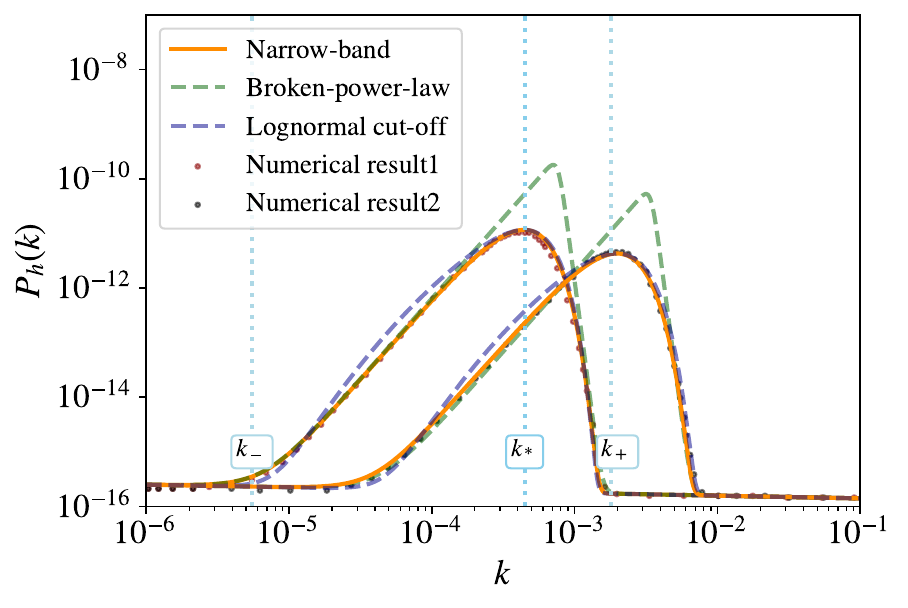}
 \includegraphics[width = 240pt]{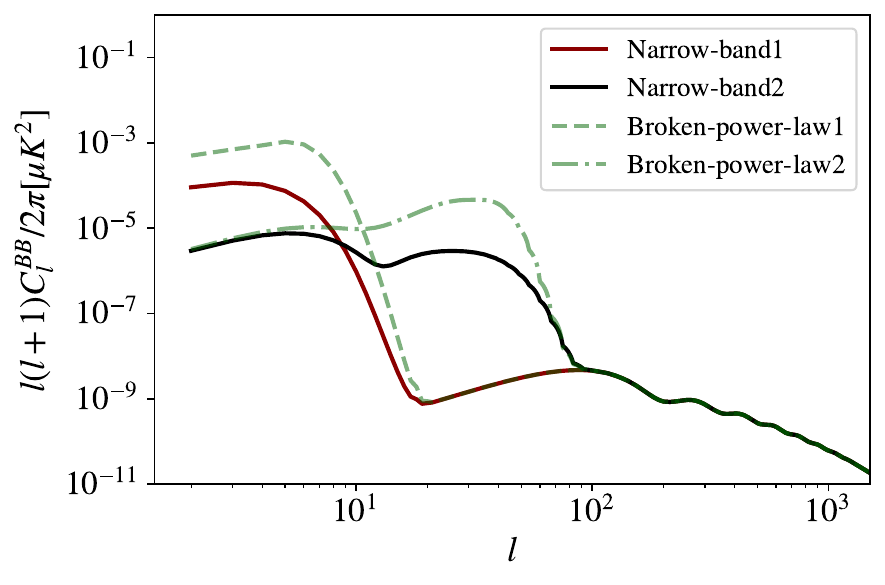}
 
\caption{\label{dotsolid of parametrization}The upper panel illustrates the comparison between the numerical results obtained from \cite{PRL} and the three distinct parametrizations: narrow-band \eqref{para}, broken-power-law \eqref{broken}, and log-normal cutoff \eqref{quadcut}. Assuming negligible curvature perturbation amplification along with the resonance, we assert that our narrow-band parametrization outperforms the others in this case. 
The broken power law introduces an extra spike when the peak is narrow, and generates a non-negligible error in the BB-mode CMB anisotropy, as demonstrated in the lower panel. 
The log-normal cut-off regime encounters challenges in performing effectively in the MCMC process. 
Blue dashed lines approximately mark the characteristic zero points of \(f(k=\exp(x))=1\), which we utilize for the prior setting in the MCMC process. 
The parameter settings are as follows. The slow-roll part \eqref{combine}: \(A_sr=2.5e-16\), \(k_\text{pivot}=1e-6\), \(n_t=-0.05\). 
narrow-band parametrization: \(h_R=10^{4.8},g=2,k_*=10^{-3.35}\) for red dots and \(h_R=10^{4.3},g=1.9,k_*=10^{-2.8}\)  for black dots. 
Broken power-law: \(A=10^{6},\alpha=3,\beta=20,k_*=10^{-3.15}\) and \(A=10^{5.5},\alpha=3,\beta=20,k_*=10^{-2.5}\).
Log-normal cut-off: \(h_R=10^{4.8},g=1.35,k^*=-3.35,\Delta=1.06\) and \(h_R=10^{4.4},g=1.2,k^*=-2.7,\Delta=1.04\).
}
\end{figure}

Our parameterization can be directly applied to this model. 
Theoretically, by simply taking the limit \( k \rightarrow 0 \) namely \(k\ll p\) on \eqref{Ph}, the integral is proportional to \(k^3\); in combination with the fact that it's a narrow-band resonance amplification, our parametrization is validated in this level. 
Particularly, the upper panel of Fig.~\ref{dotsolid of parametrization} presents the numerical results for this resonance model and the corresponding parametrizations for two representative parameter sets. We illustrate the advantage of our model by comparing its performance with previously mentioned parametrizations.

Broken power-law parameterization takes the form: 
\begin{equation}
f_\text{BPL}(k)=A \frac{\alpha+\beta}{\beta (k/k_*)^{-\alpha}+\alpha(k/k_*)^\beta} ~ ,
\label{broken}
\end{equation}
where \(\alpha,~\beta> 0 \) describe respectively the growth and decay of the spectrum around the peak, and \(k^*\) the position of the peak of the spectrum.
 Clearly it does not fit well with the narrow-band case. Given that \(\alpha\) is fixed at 3 for the \(k^3\) IR scaling, \(\beta\) needs to be large in order to match the cut-off character. This results in an unnatural peak with an acute edge and imprecisely higher amplitude, which eventually leads to error in CMB signal up to magnitudes, as shown in the lower panel of Fig. \eqref{dotsolid of parametrization}.

We also examined log-normal parametrization with a cutoff. To our convenience we use the following form
\begin{equation}
f_{LNC}(k)=h_R \exp \left(-\frac{\log ^2\left(\frac{k}{k_*} \exp \left(\left(\frac{k}{k_*}\right)^g-1\right)\right)}{2 \Delta ^2}\right)
\label{quadcut}
~ , \end{equation}
so that it matches the numerical curve in an acceptable way. However, log-normal function means a quadratic rather than linear dependence on \(\ln k\) in the component, which introduces one more parameter and consequently a degeneracy (between \(\Delta\) and \(g\) in this case). That obstructs MCMC from constraining parameters effectively. 

It is also feasible to draw a connection from the underlying model parameters to the parametrization. As to the resonance model, the most amplified wavenumber \(k_*\) for the field perturbation \(\delta\phi_k\) was approximated as the mode that exits horizon at the same time when the background stops oscillating \cite{Zhou}.
\begin{equation}k_*=Ha_I\exp(H \phi^{-1}(\phi_e))~ , \end{equation}
where \(a_I=a(t=0)\) is the scale factor at the inception of resonance, and \(\phi_e\) signifies where \(V(\phi)\) changes behavior and no longer induce oscillation and resonance. We also consider \(k_*\) as the most enhanced mode of the tensor spectrum.

Next, by evaluating \(\Delta_t^2\) in the IR limit, we find \(\Delta_t^2=bk^3\), where \(b\) is approximated by
\begin{equation}b\approx\frac{4}{15\pi^4 M_p^4}\int dl l^6\left|\int_{\tau_i}^{\tau}d\tau_1 \frac{(\tau-\tau_1)^3}{3\tau_1^2}\sum_{I}\phi_{I}\left(l, \tau_{1}\right)^2\right|^2 ~ . \end{equation}\label{bapp}
From this, we understand how to approximate \(h_E\) as mentioned above
\begin{equation}h_E\equiv h_R*r\approx b*k_*^3~ . \end{equation}

However, translating the constraints on \(h_E,k_*\) back onto the underlying model proves to be challenging. Model parameters or even model mechanisms degenerate based on observational evidence. Even if the predicted feature is observed, the choice of model still comes with various possibilities. 
In other words, once the observational feature can be depicted by less degree of freedom as in this case, a constraint on the underlying model is meaningless. Nevertheless, the ambiguity of models underscores the necessity of the parametrization we introduced, which is the main focus of this work.

\begin{figure}
\centering
\includegraphics[width = 240pt]{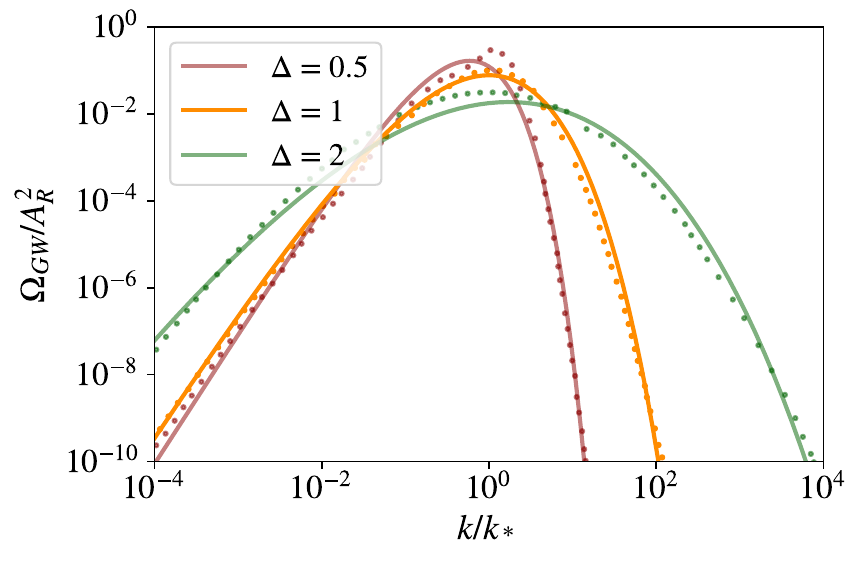}
\caption{\label{SIGW}The parametrization of the SIGW spectrum from \cite{pi2020}. The dotted line represents the original SIGW spectra, while the solid lines represent our parametrizations. From a practical perspective, we leave the parameter \(k_*\) free from the value fixed by original spectra when carrying out parametrization fitting. }
\end{figure}

Except for the two-field resonance, our parameterization is particularly effective for narrow-band GW production. One such application is scalar-induced gravitational waves (SIGWs)
\footnote{Ref. \cite{parametric} provided a specific example of a phenomenon where the amplification of finite-width, narrow-band curvature perturbations induces GWs.}. 
When the power spectrum of the curvature perturbation has a (log-normal) peak with finite width \cite{pi2020}, the resulting induced GW spectra could be well parameterized by our setup, shown in Fig. \eqref{SIGW}. However, when dealing with a zero-width delta function curvature perturbation, the IR scaling of the induced GW changes from \(\propto k^3\) to \(\propto k^2\) \cite{Infrared}. 
As illustrated in Fig.\eqref{SIGW}, for \(\Delta = 1\) our parametrization works the best, where \(\Delta\) is the width of the curvature perturbation peak. For smaller \(\Delta\) the second spike near the peak begins to form, and the scaling will finally deviate from \(k^3\). For larger \(\Delta\), conversely, we observe a deviation from narrow-band amplification, then \(g\) becomes relatively smaller, making it challenging to prevent the exponential term from affecting the linear term.

\section{\label{Sec4}MCMC results}
\begin{figure}
\centering
\includegraphics[width = 240pt]{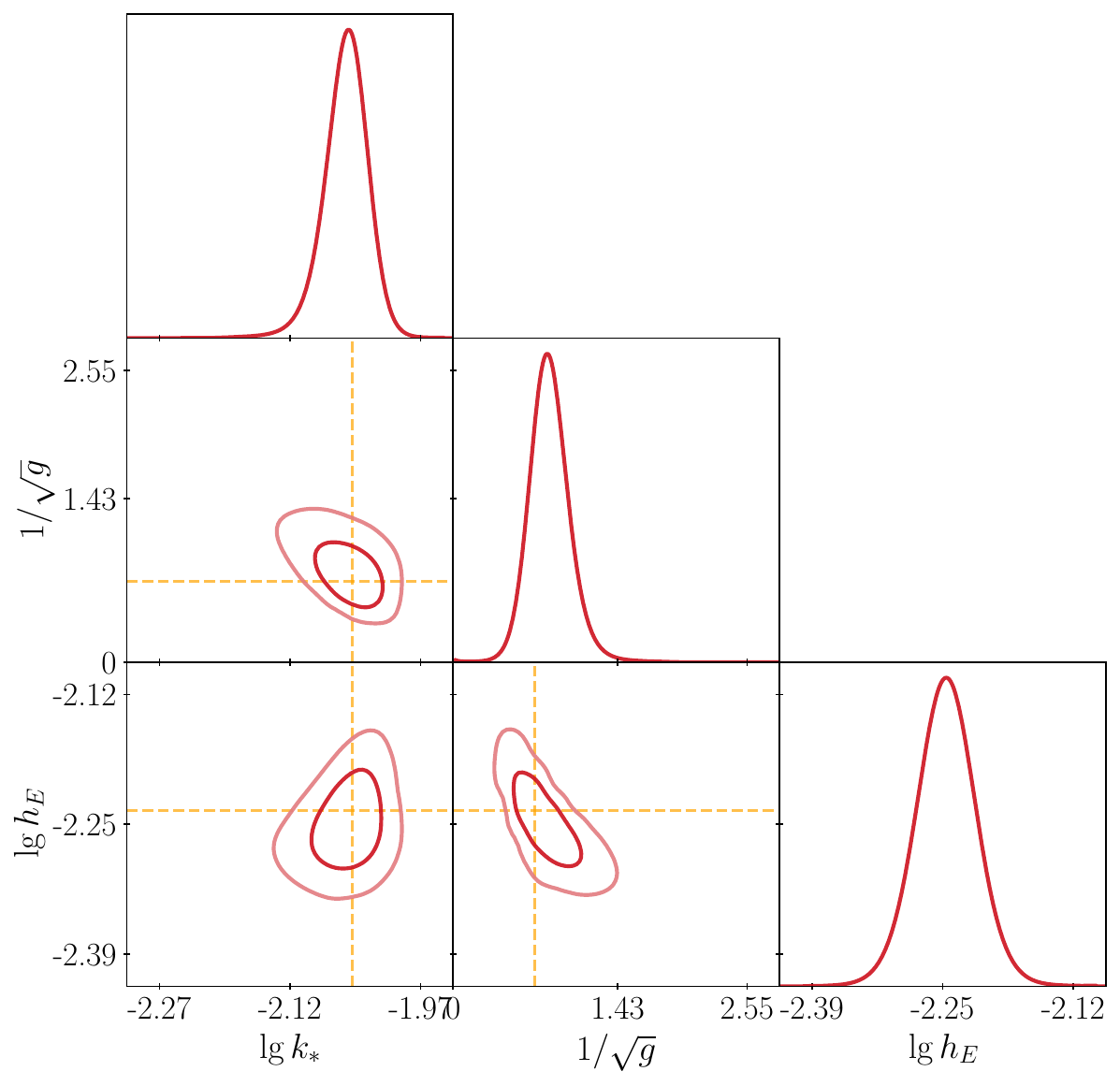}
\caption{\label{MCMC result}MCMC results using CMB-S4 mock data. Orange lines represent the fiducial values of parameters. Dark and light red denote the 1\(\sigma\) and 2\(\sigma\) confidence intervals. The adopted prior setting and assumptions are detailed in the main text. We fix the standard cosmology parameters \(\omega_b\), \(\omega_{cdm}\), \(n_s\), \(A_s\), \(h\), \(\tau_{reio}\) according to Planck results, given our assumption of an unchanged prediction on the primordial scalar power spectrum, as is the case for our primary example.}
\end{figure}

We employ the MCMC method for our parameterized model, introducing 3 additional parameters \(h_R,g,k_*\) and designating \(h_E\) as a derived parameter. To validate the feasibility of the parametrization in the MCMC process, we aim to forecast the constraining power on these parameters using mock data of CMB unlensed BB-mode anisotropies. We utilize CMB-S4 mock data alongside MontePython \cite{MontePython2018,Audren2013} and Class \cite{CLASS}, displayed in Fig.\eqref{MCMC result}.

The prior setting is derived from several reasonable assumptions as follows:

\begin{foo}
\def\labelenumi{\arabic{enumi}.}
To distinguish the scale-dependent characteristic from the power-law case, we place both the fiducial \(k_*\) and the sampled \(k_*\) within the solution domain of CLASS. 
For large values of \(g\) in \(\eqref{para}\), the damping on the UV side becomes rapid enough so that insensitive to variation of \(g\). Thus, we limit our parameter region of interest by applying \(|\ln (k_-/k_*)/\log (k_+/k_*)|\le C\). Here, \(k_-,\ k_+\) are the two solutions of \(f(k)=1\), shown in Fig.\eqref{dotsolid of parametrization}, and \(C\) is a manually chosen factor, set to be 100 here. In terms of the varying parameters, this relationship is translated to \(g\le 6 C^2/\ln h_R\). 
Given that the scalar mode remains unchanged when resonance exists, we assume the results yielded by anisotropy modes other than BB from Planck\cite{Planck18} remain veracious. Therefore, we fix \(\omega_b\), \(\omega_{cdm}\), \(n_s\), \(A_s\), \(h\), \(\tau_{reio}\) to their best-fit values, which significantly expedites the MCMC process.
\end{foo}

We set varying parameters in the form \(\log_{10} h_R\), \(1/\sqrt{g}\), \(\log_{10}k_*\), and \(\log_{10} r\), that ensures \(\Delta_t^2\) depends on them in a simple way so that MCMC yields Gaussian contours. 
For instance, the relationship \(\log k_+\approx \sqrt{2\ln h_R/(3g)}\) is a crucial character of \(\Delta_t^2\) and thus of \(C_l^{BB}\), hence we adopt \(\sqrt{1/g}\) as a parameter. Additionally, as mentioned before \(h_E = h_R*r\) rather than \(h_R\) itself is the physical parameter of the amplitude of the peak, both for the observation and the inducing mechanism. Therefore, the MCMC constraint cannot be applied to \(h_R\). Our results are presented in Fig. \(\eqref{MCMC result}\). The fiducial parameters are represented by orange lines, and the results demonstrate that our parametrization can be well constrained from observation.

\section{\label{Sec5}Conclusions}
In this article, we present a parameterization of the narrow-band primordial tensor power spectrum, as well as the gravitational wave power spectrum. Its feasibility on narrow-band GW production arises from the applicability of IR \(k^3\) scaling for transient and spiky GW sources, and the necessity of a UV side cutoff. Our method shows distinct advantages over commonly used parametrizations, particularly when applied to narrow-band amplified GWs.
We illustrate this mainly by using the specific example of a two-field inflation model that resonantly amplifies GWs.

We employ the MCMC process intertwined with CMB-Stage4 mock data to demonstrate the suitability of this parametrization for the MCMC analysis of CMB anisotropies, thereby enabling us to forecast the sensitivity on its parameters. Furthermore, by our construction one can effectively extract information from the observations and MCMC process, aiding us in further exploring the parameter range of the underlying model. 

One of the extensions of the parameterization refers to a different slope on IR scaling. Ref. \cite{Infrared} presented a physical understanding that the \(k^3\) scaling of the superhorizon modes is caused by causality. 
However, when modes are subhorizon, this scaling can still have a physical explanation. There will be a k-dependence in a trigonometric function \(\cos(k(\tau_1-\tau_2))\) when we take the correlation of the source to compute \(\Omega_{\rm GW}\), where \(\tau_1, \tau_2\) represents two source-existing moments, and we want this trigonometric function to be constant. One approach is that for both moments the source is outside horizon (\(k\tau_1, k\tau_2\ll 1\)). Another choice is that the source only persists for a short time, so that \(\tau_1-\tau_2\) is small enough. However, this discussion on scaling is limited to radiation-dominated or inflationary era. A different IR scaling is not excluded, for example when GW production appears in matter-dominated era. 

Another extension concerns amendments for cosmological probes other than CMB, such as pulsar timing array.  We anticipate it introduce no additional complications, given the conciseness and independence of the three parameters. The requirement for a narrow-band condition is stringent, and we anticipate its stability in the face of variations. We will conduct further research in this direction.

In the era of precision cosmology, model-independent approaches such as this parameterization remain crucial for qualitatively understanding the relationship between the theoretical model and its observational features. Therefore, it is beneficial to construct the parameterization more carefully, considering its specific characteristics.

\begin{acknowledgments}
The authors thank Elisa Ferreira, Xinchen He, Xiao-Han Ma, Toshiya Namikawa, Larissa Santos and Misao Sasaki for fruitful comments. This work is supported in part by the National Key R\&D Program of China (2021YFC2203100), CAS Young Interdisciplinary Innovation Team (JCTD-2022-20), NSFC (12261131497, 11653002), 111 Project for ``Observational and Theoretical Research on Dark Matter and Dark Energy'' (B23042), Fundamental Research Funds for Central Universities, CSC Innovation Talent Funds, USTC Fellowship for International Cooperation, USTC Research Funds of the Double First-Class Initiative. Kavli IPMU is supported by World Premier International Research Center Initiative (WPI), MEXT, Japan. JJ was supported by the National Research Foundation of Korea (NRF) grant funded by the Korea government (MSIT) (2021R1A4A5031460). 
We acknowledge the use of computing clusters LINDA \& JUDY of the particle cosmology group at USTC.
\end{acknowledgments}

\bibliography{references}

\end{document}